\documentclass{aastex631}

\usepackage{dblfloatfix}

\begin{document}

\title{Mapping Galaxy Images Across Ultraviolet,
Visible and Infrared Bands Using Generative Deep
Learning}

\author{Youssef Zaazou}
\affiliation{Department of Mathematics and Statistics,
Memorial University of Newfoundland, St.\ John’s, NL, A1C 5S7, Canada}
\email{yazaazou@mun.ca}

\author[0000-0002-4024-3352]{Alex Bihlo}
\affiliation{Department of Mathematics and Statistics, Memorial University of Newfoundland, St.\ John’s, NL, A1C 5S7, Canada}
\email{abihlo@mun.ca}

\author[0000-0002-6238-9096]{Terrence S. Tricco}
\affiliation{Department of Computer Science,
Memorial University of Newfoundland,
St.\ John’s, NL, A1B 3X5, Canada}
\email{tstricco@mun.ca}

\begin{abstract}

We demonstrate that generative deep learning can translate galaxy observations across ultraviolet, visible, and infrared photometric bands. Leveraging mock observations from the Illustris simulations, we develop and validate a supervised image-to-image model capable of performing both band interpolation and extrapolation. The resulting trained models exhibit high fidelity in generating outputs, as verified by both general image comparison metrics (MAE, SSIM, PSNR) and specialized astronomical metrics (GINI coefficient, M20). Moreover, we show that our model can be used to predict real-world observations, using data from the DECaLS survey as a case study. These findings highlight the potential of generative learning to augment astronomical datasets, enabling efficient exploration of multi-band information in regions where observations are incomplete. This work opens new pathways for optimizing mission planning, guiding high-resolution follow-ups, and enhancing our understanding of galaxy morphology and evolution.

\end{abstract}


\section{Introduction}

Galaxies can appear quite different when observed across various photometric bands due to the wavelength-dependent nature of the light they emit \citep{SFRindicators}. Each photometric band captures a specific range of wavelengths, highlighting different physical properties of galaxies \citep{galaxyMorph}. In shorter wavelength bands, like the ultraviolet ($U$) band, galaxies reveal the presence of hot, young stars and regions of active star formation. Meanwhile, in optical bands (e.g., $G$, $R$), the light primarily comes from older, cooler stars. At longer wavelengths, such as in infrared bands, the emission is dominated by dust and the cooler, more evolved stellar populations. 

Studying galaxies across multiple photometric bands is essential because each band reveals different aspects of a galaxy's physical properties, offering a more complete picture of its structure, composition, and evolution. By combining observations from various bands, it is possible to trace a galaxy's star formation history, stellar populations, dust content, and even interactions with its environment. Observations made in multiple wavelengths require fewer inferences and lead to stronger conclusions being drawn about the observed galaxies. This makes multi-band astronomy critical for developing a detailed understanding of a galaxy's life cycle and the processes driving its evolution. For further details, see \cite{SFRindicators} and \cite{galaxyMorph}.

\subsection{Image-To-Image Translation}

Image-to-image translation refers to a class of computer vision modeling techniques that aims to transform an image from one domain into another while preserving essential visual content \citep{img2img}. This process has gained popularity due to its applications in various fields such as medical imaging \citep{MedicalImageSegmentation}, artistic style transfer \citep{styleTransfer}, and even astronomical observations \citep{hubbleWebb}. The main challenge lies in learning the complex relationships between the source and target domains while maintaining consistency in features such as structure, texture, and color. Common approaches involve deep learning techniques, particularly convolutional neural networks.

One of the most prominent methods in image-to-image translation is generative adversarial networks, or GANs introduced in \cite{goodfellow2020generative}. The pix2pix framework was later developed by \cite{pix2pix} as a conditional GAN-based approach and quickly became a foundation for many advancements in this area. It leverages paired datasets of input and ground truth images to learn the mapping between the two domains. However, in scenarios where paired data is not available, unpaired approaches like CycleGAN \citep{cycleGAN} are employed. CycleGAN allows translation between two domains without paired images by introducing cycle consistency loss, which ensures that translating an image from one domain to another and then back again yields the original image. 

More recently, however, \cite{denoisingDiff} introduced denoising diffusion probabilistic models (DDPMs) which have succeeded GAN models as the state-of-the-art in image-to-image translation \citep[e.g.][]{sd3}. \cite{palette} developed a DDPM-based image-to-image model that consistently outperforms GAN-based implementations. 

\subsection{Deep Learning Applications in Astronomy}

Traditional data analysis methods in astronomy, particularly for studying galaxies, have been vastly outpaced by the enormous volume of data generated by modern telescopes and surveys, as outlined by \cite{bigData}. These traditional approaches often involved painstaking manual inspection and interpretation, making them time-consuming and less efficient. An example of this would be the Galaxy Zoo project, see \cite{gz2}, where crowdsourcing was used to assign morphological labels to images of galaxies. However, deep learning algorithms have revolutionized this process by automating many tasks once performed manually. For example, deep learning has proven instrumental in classifying galaxies by identifying their various types and structures as in \cite{galaxy_classification}. This automation has significantly accelerated the analysis process and increased reliability, allowing astronomers to handle larger datasets and uncover new insights more effectively. For a comprehensive dive into machine learning and deep learning applications in astronomy, see \cite{machine_survey}.

Initially, discriminative learning was favored over generative learning because it
was more straightforward and computationally feasible given the technology and resources available.  Discriminative models, which focus on classifying data and predicting labels by learning the boundaries between different classes, were simpler to implement and required less computational power relative to generative models. Recently, however, generative learning has become popular in astronomy due to significant advancements in computational power and the development of sophisticated algorithms. Examples of generative learning in astronomy can be seen in \cite{astrovader} and \cite{galaxy_simulation_diff}. However, these approaches are for unconditioned generative modeling where the output is not conditioned on any input. Image-to-image models are a subset of generative models that differ from their unconditioned counterparts in conditioning the output image on a particular input image. 

Image-to-image modeling has seen some applications in astrophysics in recent years. Initially, image-to-image modeling aimed at reconstructing noise signatures and denoising input images. \cite{reconstructNoise} was the first attempt to apply image-to-image techniques to reconstruct noise signatures and perform translation between sky surveys. \cite{UnetDenoise} proposed a U-net-based approach for image denoising and enhancement. Similarly, \cite{selfDenoise} presents a denoising method that differs from \cite{UnetDenoise} by requiring no labeled data and instead relying on self-supervision \citep{self2self} to train their models. More recently, \cite{hubbleWebb} have trained and compared several image-to-image models, including both GAN and DDPM based models, to translate Hubble Space Telescope (HST) data into James Webb Space Telescope (JWST) imagery. However, their approach is for indiscriminate patches of sky and does not isolate any one class of astronomical phenomena (e.g., galaxies, stars, nebulae).

\subsection{Our research}

In this paper, we explore the problem of translating across different photometric bands \citep{photometric} for Galaxy observations using a custom generative image-to-image machine learning model. The challenge of translating images of galaxies across photometric bands using image-to-image modeling remains relatively unexplored. This offers a promising area for further research and exploration. Despite the great successes of image-to-image machine learning models in computer vision, for example, CycleGAN \citep{cycleGAN}, pix2pix \citep{pix2pix}, and Palette \citep{palette}, the adoption of such methods in the astronomy and astrophysics community is significantly lagging. This is despite an abundance of data that could be used to train and validate image-to-image machine learning models to perform various tasks.

A successful image-to-image machine learning model could be utilized for a variety of applications. One option is to optimize observational mission planning. Existing data could be leveraged to simulate or predict what unexplored wavelengths or regions of the sky might look like, allowing prioritization of areas of interest that are likely to yield important discoveries. This could assist efficient allocation of telescope time, but also help in the planning stages for the next generation of telescopes. Another potential application is to help guide high-resolution follow-up observations. The idea here is that translated images could predict structures or regions of interest that are unresolved in the low-resolution data. These regions could be prioritized for further high-resolution study, saving observational time and resources.

 We present a fully supervised image-to-image machine learning model that is capable of predicting output bands using a variety of input bands. We also present two types of band translation -- the first being band interpolation where an output band is predicted based on two delineating input bands, while the second is band extrapolation where a sequence of bands is extended. We demonstrate that the same model architecture can be used to perform either type of translation, requiring no modifications when switching between approaches or when changing the target output band. To clarify, each set of inputs and corresponding target band outputs requires re-training the model, but it is the model architecture and training procedure that is identical for all input and output band combinations.

The majority of our results are obtained using mock observations from the Illustris simulations \citep{Illustris, IllustImg}. Illustris offers the advantage of having a wide range of bands to map between without having to account for instrumental differences as would be the case using inputs from different surveys or instruments. To confirm the applicability of our model on real-world data, we additionally present a mapping using real observations from the DESI Legacy Imaging Surveys, or DECaLS, see e.g., \cite{decals}. 

All the models’ outputs were investigated on two grounds: qualitative and quantitative. Through qualitative visual inspection, we confirm that the generated images are nearly identical to their respective ground truth images in structure. Several quantitative metrics are used to confirm our qualitative findings. General image comparison metrics are used, such as the Structural Similarity Index (SSIM), as seen in \cite{ssim}, and Peak Signal-to-Noise Ratio (PSNR), both of which function as measures of proximity between the generated and ground truth images. Additionally, we use Galaxy morphology indicators, specifically the GINI and M20 measurements, as presented in \cite{GINIM20}, to confirm that the generated images are indeed matching the ground truth images from a Galaxy morphology perspective.

To ensure reproducibility and foster further research, the code developed for this study is publicly available on \href{https://github.com/yazaazou/Galaxy-Band-Conversion}{GitHub}\footnote{\url{https://github.com/yazaazou/Galaxy-Band-Conversion}}.

\section{Proposed Approach}

\subsection{Dataset}  \label{sec:dataset}

Two distinct datasets are used throughout this work. The first dataset is a collection of mock observations obtained from the Illustris simulations. Most of our testing and prototyping has been conducted using the Illustris mock observations. The second data set is a collection of real observations of galaxies taken from the Galaxy10 DECaLS dataset courtesy of the Python package \texttt{astroNN} by \cite{astroNN}. This dataset provides a proof of concept to demonstrate that our models can be trained on and make inferences from real data.

\subsubsection{Illustris Dataset}

The Illustris mock observation catalog \citep{IllustImg} is a synthetic dataset generated from the Illustris cosmological simulations \citep{Illustris}, which model the formation and evolution of galaxies over cosmic time. The Illustris project simulates a wide range of physical processes that govern galaxy formation, including dark matter dynamics, gas cooling, star formation, supernova feedback, and black hole accretion. These simulations capture the complex interplay between baryonic matter and dark matter, allowing for the detailed modeling of galaxy evolution across cosmic time. Additionally, the simulations incorporate radiative transfer, enabling a more accurate representation of how light interacts with gas, dust, and stars within galaxies.

There are many unique advantages and disadvantages of using mock observations as opposed to real observations. Firstly, the images are ideal in the sense that they contain no instrumental noise or artifacts. Additionally, Illustris mock observations are not limited by factors such as atmospheric inference and survey depth. The most significant contribution that Illustris offers is its wide range of bands, spanning all the way from the far ultraviolet to the mid-infrared. This is significantly wider than any single survey can offer and allows us to test various band combinations without having to switch between different surveys and account for cross-survey differences such as differing noise signatures, resolutions, or misalignment issues.

Our dataset comprises paired galaxy images from the Illustris mock observation catalog where pairs are grouped by individual galaxies. Images are scaled using asinh scaling \citep{asinh}, followed by normalization to a range of $[0,1]$, as is standard in machine learning, see \cite{norm}. We used the $FUV$, $NUV$, $U$, $G$, $R$, $Z$, and $K$ bands in various combinations.  These bands correspond to bands from the Galaxy Evolution Explorer survey, or GALAX \citep{galax}, the Sloan Digital Sky Survey, or SDSS  \citep{sdss}, and the Two Micron All-Sky Survey, or 2MASS \citep{2mass}. See Table (\ref{tab:bands}) for a breakdown of the photometric bands used in this work.

\begin{table*}
\centering
\caption{Summary of Photometric Bands Used  \label{tab:bands}}
\begin{tabular}{c||c|c|c|c|c|c|c}
\hline
\hline
Band & $FUV$ & $NUV$ & $U$ & $G$ & $R$ & $Z$ & $K$  \\
\hline
Corresponding Telescope & GALAX & GALAX & SDSS & SDSS & SDSS & SDSS & 2MASS\\
Effective Wavelength (nm) & 151.3 & 230 &357.3 &472.4 & 620.1 & 891.7 & 2162.0 \\
\hline
\end{tabular}
\end{table*}

\subsubsection{Galaxy 10 Decals Dataset}
\label{sec:decals}

The Galaxy10 DECals dataset, as obtained through the \texttt{astroNN} Python package from \cite{astroNN}, contains over 17 thousand colored galaxy images available in $G$, $R$, and $Z$ bands. The observations are collected from the DESI Legacy Imaging Surveys or DECaLS \citep{decals}. The images are normalized to scale the image pixel values from $[0,255]$ to $[0,1]$, but otherwise the images are not altered or preprocessed in any other way.

\subsection{Metrics}

Various metrics are used to assess the quality of the output images as compared to target images. Some of these metrics are general image comparison metrics, in the sense that they can be applied to any set of input images. Others are specialized metrics specifically designed for analyzing galaxy morphology. We use the following set of metrics:

\begin{itemize}

\item MAE: The mean absolute error between output images and target images. This averages the absolute per pixel differences.

\item SSIM: The Structural Similarity Index, or SSIM \citep{ssim}, measures the distance between a pair of input images. Unlike the pointwise MAE, the SSIM can account for local structural differences. The SSIM has a range of $[-1.0,1.0]$ with values closer to 1.0 indicating high agreement between the images, 0.0 indicates no relation and values closer to -1.0 indicate disagreement between the inputs.

\item PSNR: The Peak Signal-to-Noise Ratio (PSNR) is a widely used metric for assessing the similarity between two images, particularly in the context of image compression and reconstruction. It measures the ratio between the maximum possible pixel value (signal) and the power of the noise (difference) between the original and the reconstructed image. PSNR is expressed in decibels (dB), with higher values indicating greater similarity, as they imply lower error between the images. A PSNR value around 30–50 dB is generally considered to imply that the reconstructed image is of high fidelity, with the reconstructed image almost indistinguishable from the original to the human eye for PSNR around 40 dB.

\item GINI and M20: The Gini and M20 measurements, first presented in \cite{GINIM20}, are used to characterize the morphological features of galaxies. The Gini coffecient is a statistical measure originally used to quantify income inequality, but in galaxy morphology, it quantifies the distribution of light among a galaxy’s pixels. The M20, on the other hand, is a measure of the relative contribution of the brightest 20 percent of a galaxy’s light to its overall light distribution. Specifically, it quantifies
the spatial distribution of the brightest regions within a galaxy helping to identify whether the galaxy has a single dominant core or multiple bright regions, which could
indicate complex internal structures or recent mergers. 

When used together, GINI and M20 provide a powerful tool for differentiating between various morphological types, especially in galaxies with irregular or disturbed structures. We will be comparing GINI and M20 measurements of output and target images to confirm that the outputs are free of spurious effects and reliably resemble natural images. Larger agreement between the distributions for ground truth label values and the generated image values indicate higher model performance. To measure the distance between these distributions, we will use the first Wasserstein distance ($\mathcal{W}_1$), also known as the earth movers distance \citep{earthMove}. To ensure the interpretability of the $\mathcal{W}_1$ measurements, we scaled the GINI coefficient and M20 measurements to a range of $[0.0, 1.0]$. However, the maximum and minimum values used for rescaling were selected globally to preserve the relative scale of the distributions. Due to this normalization, the $\mathcal{W}_1$ distance for our GINI and M20 distributions has values of $0.0$ for identical distributions and $1.0$ for maximally disjointed distributions.

\end{itemize}

\subsection{Inference Techniques}

Most sky surveys rarely capture information in one band. For example, in conducting translations from the 2MASS’s $K$ band observation \citep{2mass}, information from 2MASS’s $H$ band observations could be leveraged with minimal additional extraction and preprocessing overhead. To fully exploit all the information available from any given survey, we introduce band extrapolation. With extrapolation, a sequence of bands is extended to infer bands further along the sequence. For instance, we could select bands $G$ and $R$ as input and train a model to extend this sequence in either direction by inferring lower frequency bands (such as $Z$ and $K$), or higher frequency bands (such as $U$ or $NUV$).

We also introduce band interpolation, where instead of extending a sequence of observations, intermediate observations are generated at specific bands delineated by both input bands. For example, for inputs such as $U$ and $K$, we could generate $G$, $R$, and $Z$ observations. To clarify, each output band requires a new instance of the model. This leads to the training of multiple models to any combinations of input bands to different output bands. This applies to both interpolation and extrapolation approaches.

There are inherent pros and cons for each approach. The extrapolation models are at a disadvantage compared to the interpolation models, as they must extend in wavelength away from the input bands. Interpolation is able to use information from both high and low wavelength bands, by comparison. However, band interpolation will often require that the data will be gathered from multiple surveys, making it logistically more challenging.

\subsection{Generator Training}

The inputs of the model are denoted as $X_{\rm ph}$ where ${\rm ph}$ gives the photometric band such that $X_{\rm U}$ and $X_{\rm K}$ denote inputs from the $U$ and $K$ photometric bands respectively. The output of the model will be denoted by $\hat{Y}$ where $Y$ is to be substituted with the target output band. In this manner, a model accepting inputs $X_{\rm NUV}$ and $X_{\rm K}$ having outputs $\hat{R}$ is a model trained to interpolate the $R$ band from corresponding $NUV$ and $K$ inputs. In the previous example, the ground truth labels would be denoted as $R$, and this notation will carry forward in subsequent sections.

We elected to employ a fully supervised training process to train our generator models. Our models can be considered a complex non-linear transformation with the model weights controlling the transformation applied to the input(s). Training our models consists of finding the model weights that minimize the distance between the output and the ground truth label. This distance is referred to as the loss function. 

Our loss function is the combination of the $\mathcal{L}_1 $ (mean absolute) error and the inverse SSIM loss given by,
\begin{equation}
\mathcal{L}_{\text{SSIM}} = 1 - \text{SSIM}({Y}, \hat{Y}),
\end{equation}
where SSIM($Y$, $\hat{Y}$) corresponds to the SSIM between the ground truth and reconstructed images. This loss function was determined through a series of trial and error. The full objective to be minimized during training is
\begin{equation} \label{eq2}
\mathcal{L}= \mathcal{L}_1  + \lambda   \mathcal{L}_{\text{SSIM}},
\end{equation}
where $\lambda$ is a regularization parameter greater than 0. We experimented with setting $\lambda=0$, but found that setting $\lambda=1$ yielded higher quality results. Investigating the effects of $\lambda$ values greater than 1 would be of interest.

\subsection{Generator Architecture}

We employ a ResNet-like architecture \citep{resid}, primarily composed of residual blocks. This architecture emphasizes residual learning, making it easier for the model to learn identity mappings if needed. The generator itself follows the architecture of the generator networks in the CycleGAN implementation \citep{cycleGAN}. However, after performing hyperparameter tuning, we found the default hyperparameters in CycleGAN did not perform best compared to other configurations. In summary, we decreased the number of downsampling and upsampling blocks to increase the resolution of the latent residual layers along with increasing the number of residual blocks from 6 to 9. For more details on model selection and training, see Appendix \ref{appendix}. Note that we have elected not to use a diffusion-based architecture due to the probabilistic nature of the image generation process, which may yield several outputs for any one input.

\section{Results}

\subsection{Interpolation} \label{secInter}

We now investigate the performance of three models tasked with band interpolation based on the same input. The input bands are the $NUV$ and $K$ band which correspond, in wavelength, to GALAX’s $NUV$ and 2MASS’s $K$ band respectively. We note significant differences between both input bands since we have selected the bands to be quite far apart to provide the models with the opportunity to showcase their ability to handle non-trivial transformations.

We trained three models to map to $G$, $R$, and $Z$ bands corresponding in wavelength to SDSS bands. The reason for selecting those bands is to investigate the theoretical augmentation of SDSS’s library using GALAX and 2MASS observations. A paired set of observations taken from both GALAX and 2MASS could be used to augment the SDSS database significantly. The task of obtaining and preprocessing such a paired data set is not one we have investigated, but is theoretically possible and would potentially produce high-quality images to augment SDSS observations with minimal inference cost.

\begin{figure*}
    \centering
    {\includegraphics[trim={3.8cm 12.5cm 3cm 13.2cm},clip,width=15.5cm]{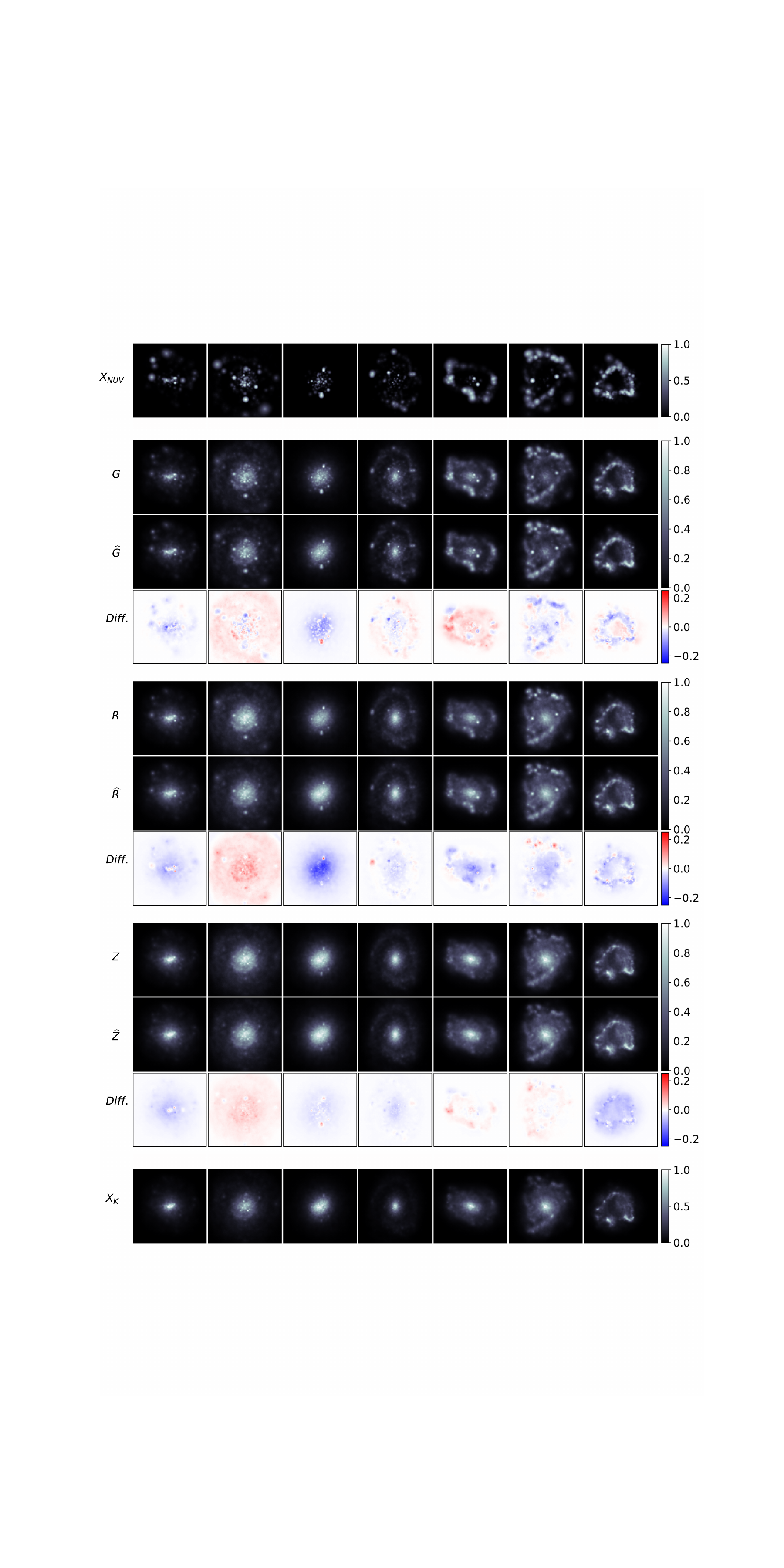} }
    \caption{The results for our three trained interpolation models; each column is reserved for a single galaxy. The $NUV$ inputs and the $K$ inputs ($X_{NUV}$ and $X_{K}$) are delineated by the $G$, $R$, and $Z$ groupings (one for each model).The first row of each grouping shows the ground truth labels, the second shows the respective generated images while the last row is the residual taken to be the ground truth label minus the model output.}
    \label{fig:inpainting}
\end{figure*}

The inputs and the interpolated results can be seen in Figure (\ref{fig:inpainting}). The generated outputs are nearly identical to the ground truth data. This is due to the model being able to take advantage of the information present in both bands. The model can detect areas of high star formation from the $NUV$ band while also having access to the distribution of the majority of the stellar matter through the $K$ band. Leveraging information from both bands allows it to make predictions that are more precise than is possible with only one band’s input.

\begin{deluxetable}{cccccc} 
\tablecaption{Performance metrics for the three interpolation models. \label{tab:interpolation_metrics}}
\tablehead{
    \colhead{Inferred Band ($\lambda_{\text{eff}}$)} & \colhead{MAE $\downarrow$} & \colhead{SSIM $\uparrow$} & 
    \colhead{PSNR $\uparrow$} & 
    \colhead{GINI : $\mathcal{W}_1$ $\downarrow$} & \colhead{M20 : $\mathcal{W}_1$ $\downarrow$} 
}
\startdata
$G$ (472 nm) & 0.0078 & 0.984 &35.3& 0.0061 &0.0064\\
$R$ (620 nm) & 0.0086 & 0.986 &34.3& 0.0206 &0.0060\\
$Z$ (891 nm) & 0.0066 & 0.990 &37.4 &0.038 &0.0098
\enddata
\end{deluxetable}

Table~(\ref{tab:interpolation_metrics}) shows performance metrics for our three interpolation models. All the shown metrics indicate that the models consistently produce output that is in line with the ground truth data. The MAE is less than 1\%. SSIM scores of ${\sim}0.99$ indicate that the generated and ground truth images are virtually identical, indicating excellent similarity in image quality assessment. For perspective, the SSIM between the $NUV$ input band and the ground truth labels of bands $G$, $R$ and $Z$ are 0.53, 0.45, and 0.43, respectively, while the SSIM between the $K$ input band and ground truth bands $G$, $R$, $Z$ are 0.86, 0.82, and 0.85, respectively.

The mean PSNR measurements are ${\sim}35$ dB, which generally implies that the generated images are of high quality. Departures from the ground truth data are small and only noticeable upon close inspection.

\begin{figure*}
    \centering
    {\includegraphics[trim={2.5cm 1.5cm 3.5cm 2cm},clip,width=18cm]{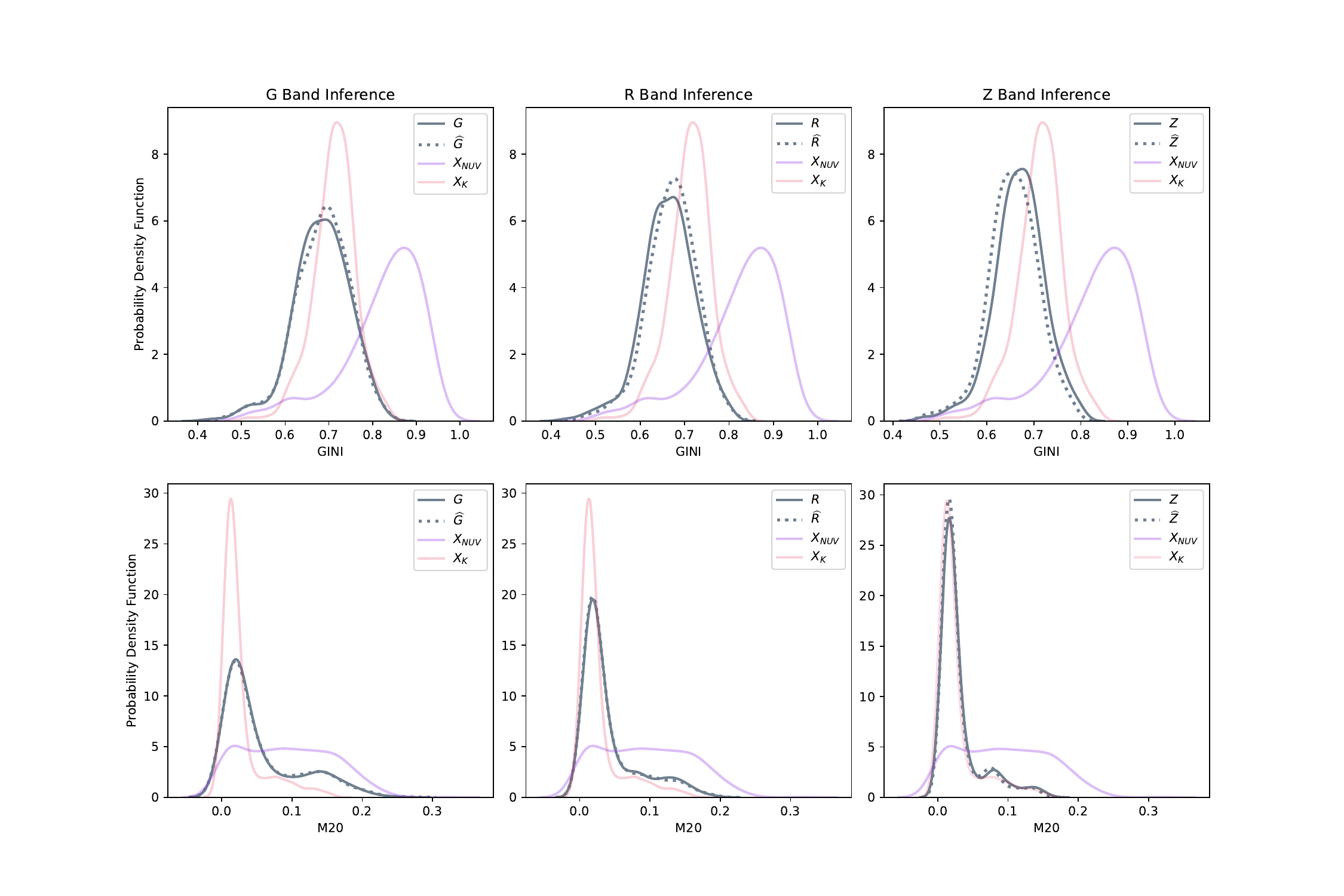} }
    \caption{This figure shows the distributions of the GINI and M20 values for each of the three interpolation models alongside the input data's distributions for reference. We note significant agreement in the ground truth and generated distributions particularly for the M20 measurement. While the GINI coefficient distributions are not perfectly aligned, they still indicate a successful transformation especially when taken in the context of the initial distributions of the inputs.}
    \label{fig:inpaint_stats}
\end{figure*}

Figure (\ref{fig:inpaint_stats}) shows the distributions of the GINI and M20 values for the input and generated data. The distributions for the input bands are represented, along with the distributions of the ground truth and generated images in the target band. The shape of the distributions between the generated and ground truth images for the target band are in excellent agreement. This highlights that the model is capturing the breadth of diversity within galaxies for the target band, avoiding issues such as mode collapse. Table (\ref{tab:interpolation_metrics}) quantifies the GINI and M20 measurements by comparing the Wasserstein distance between the two distributions. Relatively low $\mathcal{W}_1$ values are observed, indicating a significant agreement between the ground truth data and generated images, as far as Galaxy morphology is concerned.

\subsection{Extrapolation} \label{secExtra}

This section investigates the performance of three models tasked with band extrapolation based on similar inputs. The input bands are the $G$ and $R$ bands, which correspond, in wavelength, to SDSS’s $G$ and $R$ bands. The two input bands are proximal in wavelength but still provide the model with sufficient information as to the progression of band observations. Further experimentation is needed to investigate the effect of the input bands’ proximity. A different band combination (such as $G$ and $Z$) may provide the model with better information.

We trained three models to map to $U$, $NUV$, and $FUV$ bands corresponding in wavelength to the SDSS $U$ band and the GALAX $NUV$ and $FUV$ bands, respectively. All three output bands are lower in wavelength than the input bands, which presents the models with more challenging transformations. Lower wavelength inferences are more demanding than their higher wavelength counterparts because of the inherent complexity in regions containing high star formation rates. Compared to band interpolation, this task is, practically, more straightforward since it is highly probable that all of the input data can be obtained from a single survey.

\begin{figure*}
    \centering
    {\includegraphics[trim={3cm 4.5cm 2.7cm 7.4cm},clip,width=16cm]{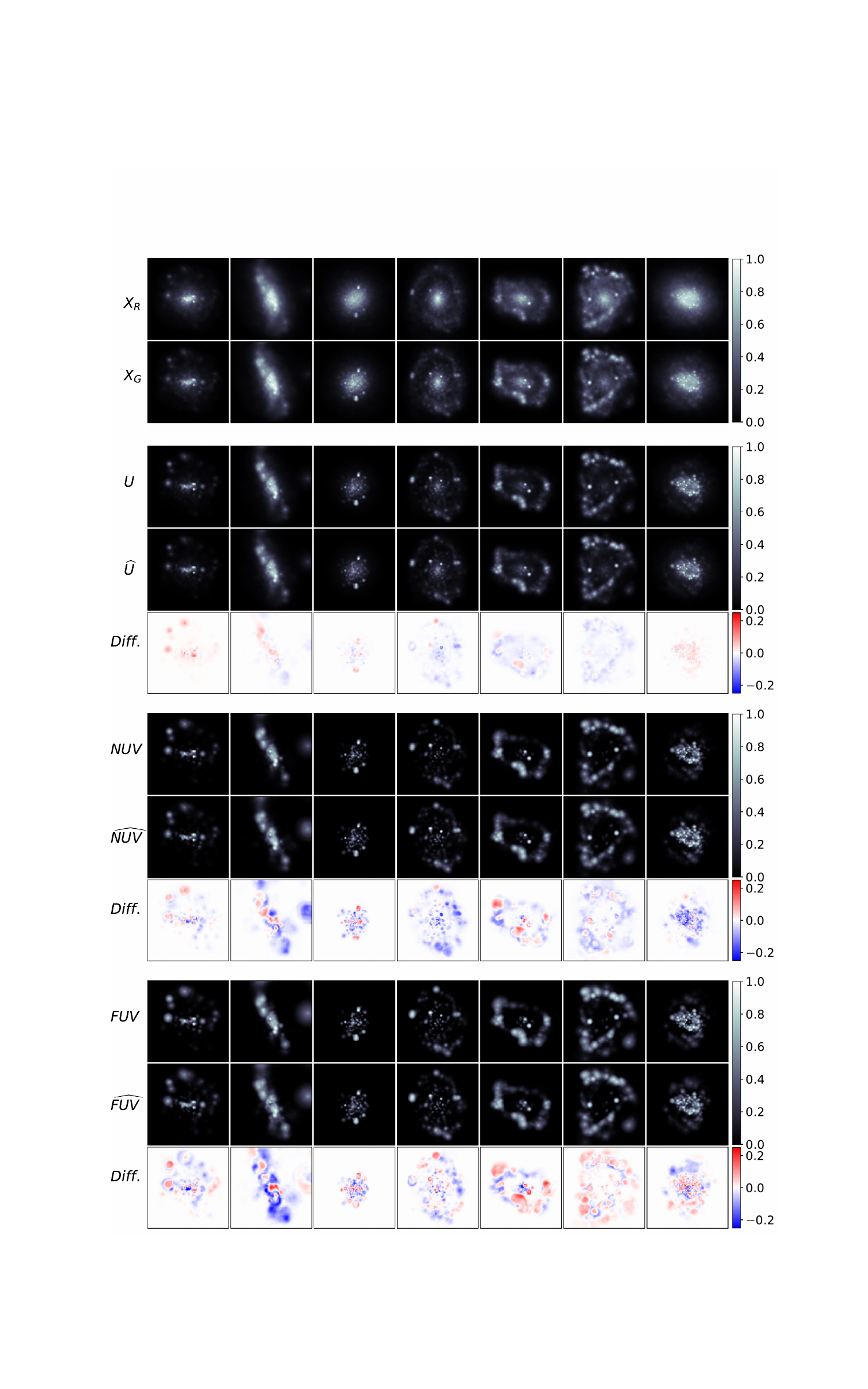} }
    \caption{The results for our three trained extrapolation models; each column is reserved for a single galaxy. The $R$ inputs and the $G$ inputs ($X_{R}$ and $X_{G}$) precede the $U$, $NUV$, and $FUV$ groupings (one for each model). The first row of each grouping shows the ground truth labels, the second shows the respective generated images while the last row is the residual taken to be the ground truth label minus the model output.}
    \label{fig:extrapolation}
\end{figure*}

The inputs and the extrapolated results can be seen in Figure (\ref{fig:extrapolation}). As with the band interpolation, the generated images are nearly identical to the ground truth data. The model is able to recreate structure present in the ground truth data with a very high degree of accuracy. Of particular note is the models' ability to preserve their accuracy across the sequence even as the output band becomes further removed from the input bands, such as in the $FUV$ band inferences, though it can be noticed that the magnitude of residuals increase as the extrapolation distance increases.

\begin{deluxetable}{cccccc} 
\tablecaption{Performance metrics for the three extrapolation models. \label{tab:extrapolation_metrics}}
\tablehead{
    \colhead{Inferred Band ($\lambda_{\text{eff}}$)} & \colhead{MAE $\downarrow$} & \colhead{SSIM $\uparrow$} & 
    \colhead{PSNR $\uparrow$} & 
    \colhead{GINI : $\mathcal{W}_1$ $\downarrow$} & \colhead{M20 : $\mathcal{W}_1$ $\downarrow$} 
}
\startdata
$U$ (357 nm) & 0.0034 & 0.99 &40.3 & 0.0086 &0.0079\\
$NUV$ (230 nm) & 0.0089 & 0.95 &32.3 & 0.035 &0.0081\\
$FUV$ (151 nm) & 0.0127 & 0.91 &28.4 & 0.049 &0.014
\enddata
\end{deluxetable}

The performance metrics in Table (\ref{tab:extrapolation_metrics}) indicate that the models consistently produce output that is in line with the ground truth data across multiple extrapolated bands. The MAE scores are ${\sim}$1\% for the three bands, with increasing error the further the extrapolated band is from the source bands.

The $U$ and $NUV$ models have SSIM scores of 0.99 and 0.95, respectively. In general, SSIM scores $>0.95$ indicate that two images are virtually identical, meaning that the images generated by our models are exceptionally representative of the ground truth images. While the $FUV$ model's SSIM score of $0.91$ indicates slightly lower agreement, this is still a perfectly acceptable result, particularly given how distant the $FUV$ band is from the input. For perspective, the SSIM between the $G$ input band and the ground truth labels of bands $U$, $NUV$, and $FUV$ are 0.84, 0.52, and 0.49, respectively, while the SSIM between the $R$ input band and ground truth bands $U$, $NUV$, and $FUV$ are 0.73, 0.44, and 0.41, respectively. The generated images are indeed high-quality transformations of the input bands.

A similar pattern is noticeable along the PSNR measurements of the three models. The $U$ model’s PSNR of above $40$ dB indicates nearly perfect reconstruction of the ground truth data. The $FUV$ and $NUV$ models average around $30$ dB, which indicates visible, but minor, differences between the generated and ground truth images.

\begin{figure*}
    \centering
    {\includegraphics[trim={2.5cm 1.5cm 3.5cm 2cm},clip,width=18cm]{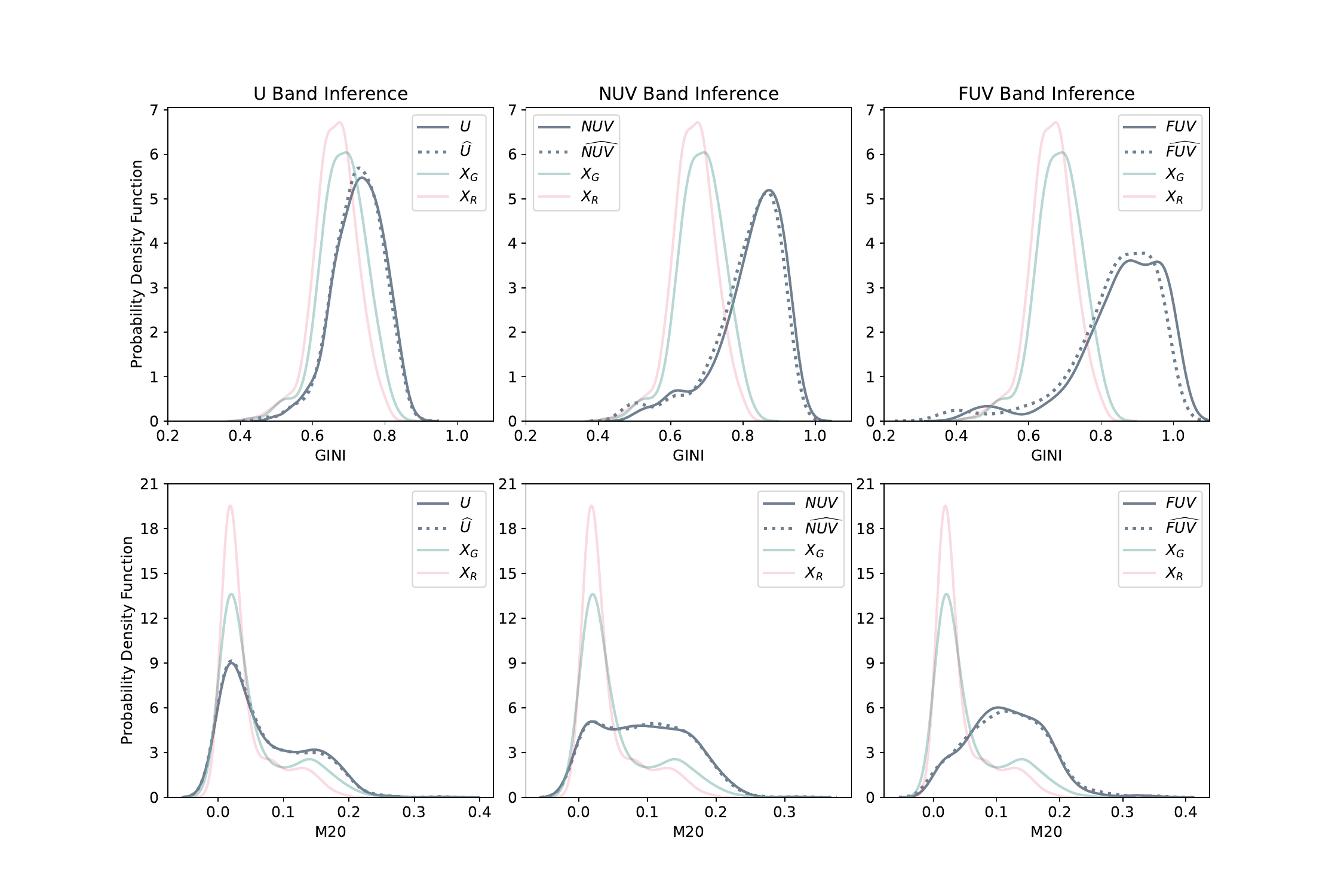} }
    \caption{This figure shows the distributions of the GINI and M20 values for each of the three extrapolation models alongside the input data's distributions for reference. We note significant agreement in the ground truth and generated distributions particularly for the M20 measurement. While the GINI coefficient distributions are not perfectly aligned, they still indicate a successful transformation especially when taken in the context of the initial distributions of the inputs.}
    \label{fig:extra_stats}
\end{figure*}

Figure (\ref{fig:extra_stats}) shows the distributions of the GINI and M20 values for our data, along with the input bands' distributions. The models do an excellent job of reproducing the distributions for the target band. Of note is the M20 values for the $FUV$ band, as the input bands are bimodal and highly overlap, whereas the $FUV$ target band is uni-modal, but non-Gaussian, with little overlap to the input band distributions. The GINI and M20 distributions for the generated images match the ground truth distributions almost perfectly across all three extrapolated bands. Table (\ref{tab:extrapolation_metrics}) shows the $\mathcal{W}_1$ distances between the generated and ground truth distributions, indicating high agreement, reflected by the relatively low $\mathcal{W}_1$ values. This provides strong evidence that the models produce representative galaxy morphologies across the data set.

\subsection{DECaLS Observations}

We train our model architecture on real observational data from DECaLS (see the data description in Section~\ref{sec:decals}). A single model is trained to interpolate the $R$ band based on input $G$ and $Z$ bands, respectively. Training on DECaLS data is highly significant, as it represents our model's capacity to learn and generate from real data, rather than mock observational data from the Illustris simulations. While the wavelength difference between the input bands and output band is smaller than the results of the models presented in Section~(\ref{secInter}), this concession is necessary to gauge how noise and various background artifacts in real observational data can complicate the training process.

\begin{figure*}
    \centering
    {\includegraphics[trim={7.8cm 10.5cm 7cm 11cm},clip,width=18cm]{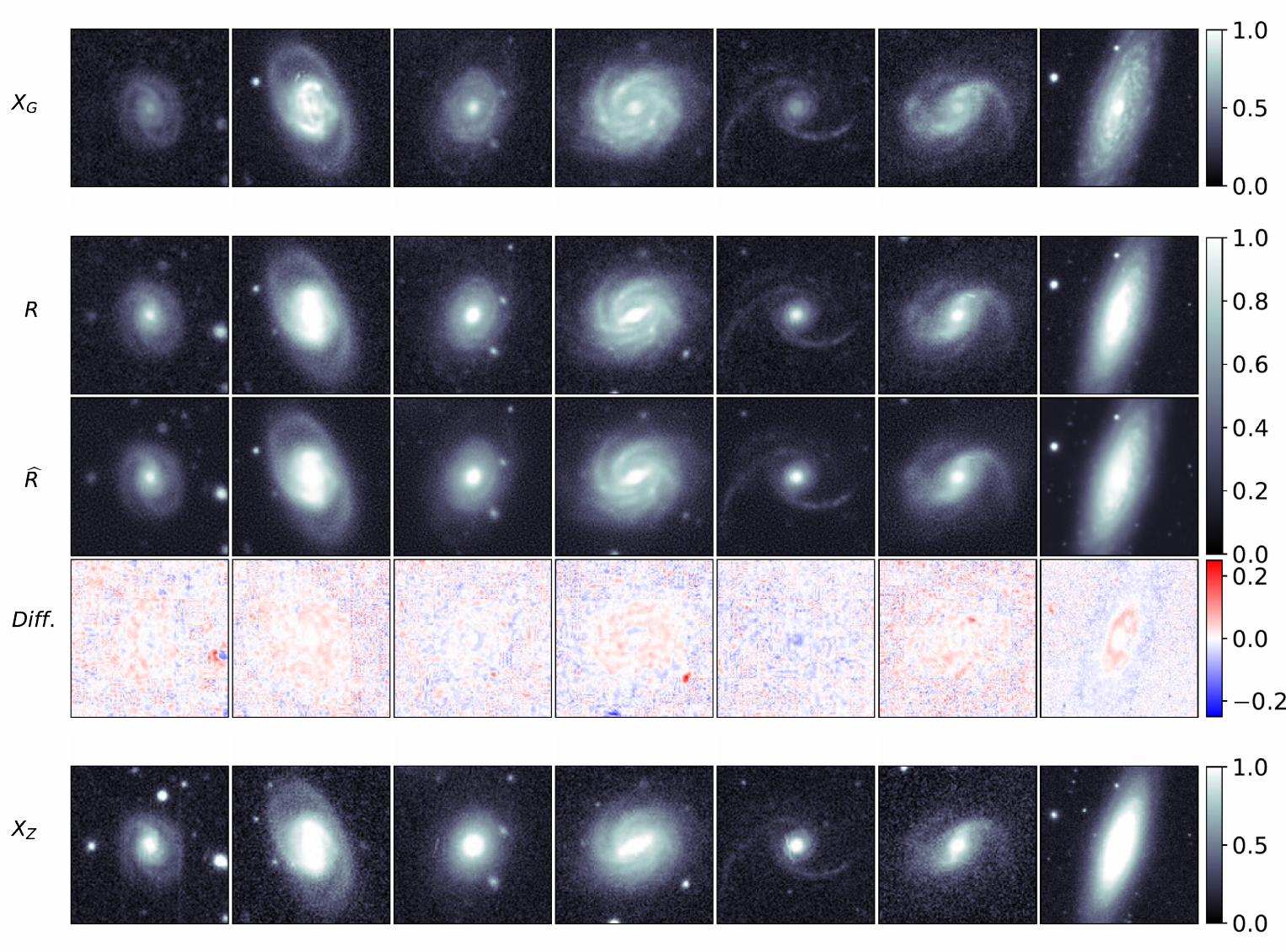} }
    \caption{The results for our DECaLS model; each column is reserved for a single galaxy. The $G$ and $Z$ inputs ($X_{G}$ and $X_{Z}$) are delineated by the $R$ grouping the first row of which shows the ground truth labels, the second shows the respective generated images while the last row is the residual taken to be the ground truth label minus the model output.}
    \label{fig:decals_inter}
\end{figure*}

Figure (\ref{fig:decals_inter}) shows the input images ($G$ and $Z$ bands), the ground truth and generated $R$ band images, along with the residuals between the generated and ground truth images. A significant level of agreement between the ground truth data and the generated images is evident. Of note is that the signatures of the central bulge are much more similar to the ground truth images compared to either of the inputs. Similarly, the outline of the spiral arms is in line with the ground truth images. The residuals contain mostly noise and are free from any significant signs of structure. The model is recreating the galaxy structure present in the ground truth data.

\begin{deluxetable}{cccccc} 
\tablecaption{Performance metrics for the DECaLS trained interpolation model. \label{tab:decals_metrics}}
\tablehead{
    \colhead{Inferred Band ($\lambda_{\text{eff}}$)} & \colhead{MAE $\downarrow$} & \colhead{SSIM $\uparrow$} & 
    \colhead{PSNR $\uparrow$} & 
    \colhead{GINI : $\mathcal{W}_1$ $\downarrow$} & \colhead{M20 : $\mathcal{W}_1$ $\downarrow$} 
}
\startdata
$R$ (620 nm) & 0.022 & 0.86 &30.5& 0.034 &0.0012
\enddata
\end{deluxetable}

Table (\ref{tab:decals_metrics}) shows our performance metrics. The MAE is ${\sim}$2\%, which is higher than the Illustris-trained models. This is evident from the residuals in Figure~\ref{fig:decals_inter}, which show a general level of background noise that is not present to such a degree in the clean Illustris data. The MAE metric will average this background noise into a non-zero value, even though this noise is largely random.

The SSIM score of $0.86$ is not as strong of an agreement as compared to the results of Section (\ref{secInter}). However, the majority of the discrepancy is between the differing noise signatures and backgrounds of the generated and ground truth images. For perspective, the SSIM between the $R$ labels and the $G$ and $Z$ inputs is 0.53 and 0.38, respectively. An increase to SSIM values of $0.86$ from either input certainly indicates that the model has learned a correct transformation to the target band. 

The PSNR score of ${\sim}$30 is similar to the results found when interpolating Illustris mock observational data, as given in Table~(\ref{tab:interpolation_metrics}). This implies that the generated images are representative of the ground truth data.

\begin{figure*}
    \centering
    {\includegraphics[trim={3.5cm 0cm 3.6cm 0.7cm},clip,width=17cm]{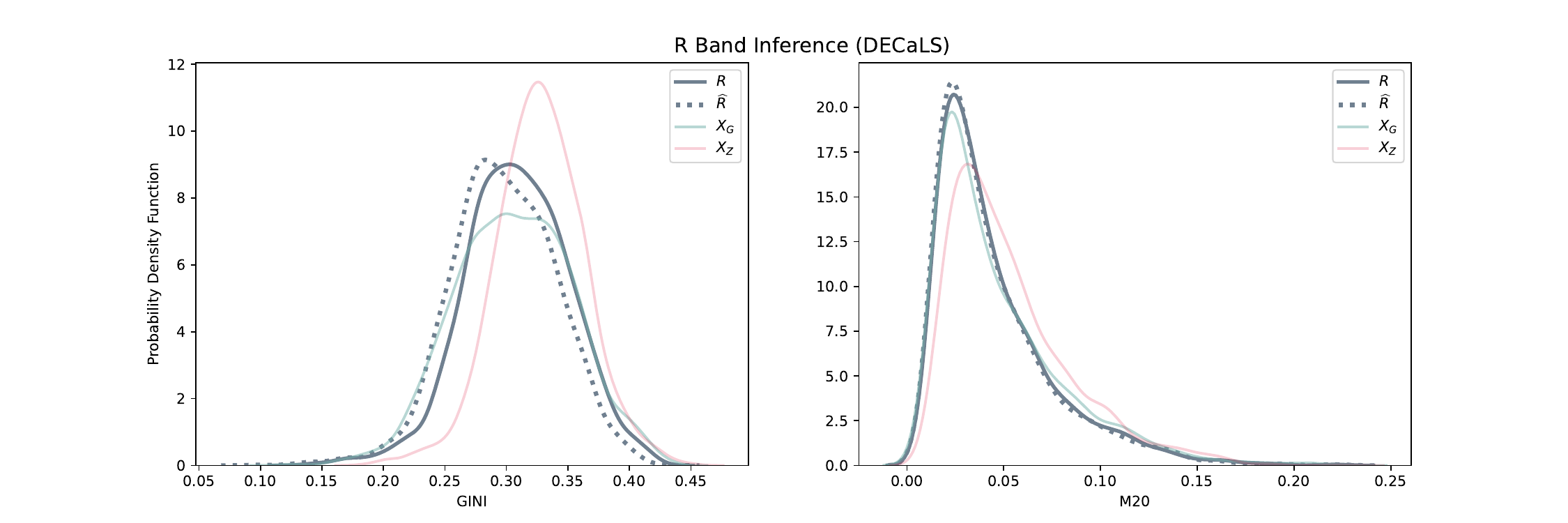} }
    \caption{This figure shows the distributions of the GINI and M20 values for the DECaLS interpolation model with the input distributions for reference.}
    \label{fig:decals_stats}
\end{figure*}

Figure (\ref{fig:decals_stats}) shows the distributions of the GINI and M20 values for the input bands ($G$ and $Z$), and the generated and ground truth bands ($R$). The model is able to reproduce the target distribution, though with some minor differences in the GINI distribution. As given in Table~(\ref{tab:interpolation_metrics}), the $\mathcal{W}_1$ distances between the distributions are found to be quite small, with similar magnitude to the $\mathcal{W}_1$ distances for the band interpolation models trained on Illustris mock observational data. As with the other metrics explored, this indicates high agreement between ground truth data and generated images in terms of galaxy morphology.

\section{Conclusion}

In this study, we have demonstrated the feasibility and effectiveness of generative deep learning models for translating galaxy observations across photometric bands. Our model is a convolutional neural network with a ResNet-like architecture that uses a combination of MAE loss and SSIM score as a custom loss function. The model is a supervised image-to-image model that can perform interpolation or extrapolation of photometric bands from input band sources. Our experiments use mock galaxy observations from Illustris simulation data and real-world observational data from DECaLS.

The main contributions of this paper include: (i) outlining a straightforward image-to-image model that relies on supervised training using a custom training loss; (ii) demonstrating that our image-to-image model successfully performs a variety of mappings between different photometric bands; and (iii) demonstrating that the Illustris mock observation catalog can be used to develop and prototype image-to-image models (and that successes with the Illustris dataset carry over into observed datasets). Our main contribution is a proof of concept that image-to-image models can be used to translate observations of galaxies across different photometric bands.

Experiments with Illustris-trained models performed band interpolation to $G$, $R$ and $Z$ bands using $K$ and $NUV$ bands as inputs, and band extrapolation using $G$ and $R$ as inputs to predict $U$, $NUV$ and $FUV$. Exceptional results across all metrics were achieved for both band interpolation and extrapolation. SSIM values ranged between 0.98 and 0.99 for interpolation and above 0.91 for extrapolation, indicating that the generated images were nearly identical to the ground truth. Similarly, mean PSNR values ranged from 30–40 dB, demonstrating minimal perceptible differences. Furthermore, specialized galaxy morphology metrics, such as the GINI coefficient and M20, displayed strong alignment with ground truth, evidenced by low $\mathcal{W}_{1}$ values between ground truth and generated distributions. This high performance reflects the advantages of working with noise-free, artifact-free data in the Illustris dataset, enabling the models to effectively capture the intricate relationships between photometric bands.

The model trained on real observational data from DECaLS performed band interpolation, taking $G$ and $Z$ band data as inputs and interpolating the $R$ band. The performance of the DECaLS model was slightly reduced compared to the Illustris-data trained model, which can be attributed to the complexities of real-world observations, such as noise and artifacts in the DECaLS data. Despite this, the mean SSIM score of 0.86 and mean PSNR values of approximately 30 dB reflect a high degree of quality in the generated images. The low $\mathcal{W}_{1}$ values between GINI and M20 distributions are comparable to values obtained with the Illustris-trained models.




By utilizing mock observations from the Illustris simulations, we have shown that these models can perform both band interpolation and extrapolation with high fidelity, producing outputs that align well with ground truth images across multiple evaluation metrics. The ability of our models to generate real-world data, as evidenced by the DECaLS observations, highlights the robustness and practical applicability of our approach.

Our findings open new opportunities for augmenting astronomical datasets in regions where multi-band observations are sparse or unavailable. This can significantly aid in mission planning and enhance high-resolution follow-ups, ultimately contributing to a deeper understanding of galaxy morphology and evolution. Future work will focus on extending the model's capabilities to address limitations such as noise variability and calibration differences across surveys, as well as exploring unsupervised or semi-supervised approaches to further enhance the model’s generalization capabilities. We would also like to experiment with different model architectures, such as diffusion models.

\section*{Acknowledgments}

We would like to thank Dr.\ Anna O'Grady for generously lending us her time and insight. This research was undertaken, in part, thanks to funding from the Canada Research Chairs program and the NSERC Discovery Grant program.

\appendix
\section{Training Details} \label{appendix}

We trained our each of our models for $128$ epochs on a V100 GPU which took approximately 2 hours for each training instance. The model uses around $12$ MB of memory and, once trained, can be used to make around $1000$ inferences in about $5$ minutes (using the same V100 GPU), making it very practical to deploy and run.

\subsection{Datasets}

Following the preprocessing discussed in Section (\ref{sec:dataset}), the images in both the Illustris and DECaLS datasets were resized to a resolution of $128 \times 128$. The Illustris training dataset is comprised of $2000$ samples (a sample is a set inputs along with an associated ground truth label), while the validation dataset contains $500$ samples. The DECaLS training dataset contains $8000$ samples while the validation dataset contains $2000$ samples.

The reason for the DECaLS datasets being larger than the Illustris datasets is that the DECaLS samples contain more noise and background artifacts which makes learning from the DECaLS dataset more challenging compared to the Illustris dataset. To compensate for this, we selected a larger number of samples for the DECaLS dataset compared to the Illustris dataset. 

\subsection{Generator Hyperparameters}

We have based our model architecture on the model presented in  \cite{cycleGAN}, which  proposed a generator architecture with a ResNet backbone as in \cite{resid}. The architecture is composed of down-sampling blocks followed by residual blocks, first presented in \cite{resid}, and finally up-sampling blocks. The down-sampling blocks reduce the spatial dimensions of the input image while capturing high-level features. This is achieved using strided convolutions, which compress the image's resolution, allowing the network to focus on broader patterns and context. The residual block preserves the input's spatial information while transforming features. It consists of convolutional layers with skip connections that add the input directly to the output, facilitating gradient flow during training. The up-sampling block restores the compressed spatial dimensions back to the original size using transposed convolutions. This process reconstructs finer details in the output image. For more information on different types of convolutions used in deep learning, see \cite{conv_types}.

\cite{cycleGAN} proposed that for images of size $128 \times 128$, the generator architecture be comprised of 2 down-sampling blocks followed by a latent space containing 6 residual blocks and 2 up-sampling blocks. We found that a different hyperparameter configuration works better for our purposes (note that hyperparameters are the configuration parameters of the model, and are not learned during training).  Our generators utilize only 1 down-sampling and up-sampling block while increasing the number of residual blocks from 6 to $9$. Our reasoning for this configuration working better is that it contains a wider bottleneck (in other words less down-sampling or compression of the input) and relies more on residual learning. A small bottleneck does not transmit high-frequency features, such as spirals and bars, but only transmits low-frequency features such as overall shape and orientation. Given that the most significant differences between photometric bands are mainly high-frequency features, it follows that
the more aggressive the bottleneck, the less adept our network is in resolving these high-frequency features.

\bibliographystyle{aasjournal}
\bibliography{template}{}

\end{document}